\documentclass[10pt,a4paper,twoside,twocolumn]{article}

\usepackage{amssymb}
\usepackage{amsmath}
\usepackage{bm}
\usepackage{graphicx}
\usepackage{subfigure}
\usepackage{color}
\usepackage[font=small,labelfont=bf]{caption}

\usepackage{booktabs}
\usepackage{array}
\usepackage{enumitem}
\usepackage[affil-it,auth-lg]{authblk}
\usepackage{eso-pic}
\usepackage{abstract}

\usepackage{natbib,twoopt}
\usepackage[]{hyperref} 
\bibpunct{(}{)}{;}{a}{}{,}

\let\OLDthebibliography\thebibliography
\renewcommand\thebibliography[1]{
	\OLDthebibliography{#1}
	\setlength{\parskip}{0pt}
	\setlength{\itemsep}{0pt plus 0.3ex}
}

\begin{document}
	
	\title{Generation of massive stellar black holes by rapid gas accretion in primordial dense clusters}
	
	\author[1]{Zacharias Roupas}
	\affil[1]{Centre for Theoretical Physics, The British University in Egypt, Sherouk City 11837, Cairo, Egypt} 
	
	\author[2]{Demosthenes Kazanas}
	\affil[2]{Astrophysics Science Division, NASA Goddard Space Flight Center, Greenbelt, MD 20771, USA} 

	\twocolumn[
\begin{@twocolumnfalse}
	\date{\vspace{-5ex}}
	\maketitle
	\begin{abstract}
        Supernova theory suggests that black holes of a stellar origin cannot attain masses in the range of 50-135 solar masses in isolation. We argue here that this mass gap is filled in by black holes that grow by gas accretion in dense stellar clusters, such as protoglobular clusters. The accretion proceeds rapidly, during the first 10 megayears of the cluster life, before the remnant gas is depleted. We predict that binaries of black holes within the mass gap can be observed by LIGO. 
		\end{abstract}
\end{@twocolumnfalse}
\hspace{10pt}
]


\section{Introduction}

Pair-instability supernova \citep{2002ApJ...567..532H} signifies the presence of an {upper mass gap at $50-135M_\odot$ of population I/II stellar origin black holes (BHs)} \citep{2016A&A...594A..97B,farmer2019mind}. {Up to the time this work is published}, no BH mass in binary black hole (BBH) gravitational wave (GW) signals detected by \cite{LIGO_2018arXiv181112907T} exceeds $50M_\odot$. 
Nevertheless, $75\%$ of these BHs have masses greater than $20M_\odot$, which is far more than expected from stellar evolution and more than is observed in X-ray binaries. 
\cite{2017ApJ...851L..25F} suggested that {because of LIGO's higher sensitivity} to BBH members $\gtrsim 20M_\odot$, an excess of higher mass BHs should be expected, with a maximum at $40M_\odot$. 

We suggest here that {besides LIGO's bias} toward higher BH masses, the BH mass function observed will favor masses that are higher than predicted by stellar evolution {because of gas accretion by BHs} in dense stellar clusters. This will occur rapidly \citep{2019A&A...621L...1R} before the gas is depleted through the first stellar formation event. More importantly, we estimate that this shift is sufficient to fill the upper mass gap. This mechanism is different, {but not mutually exclusive, from the} repeated mergers scenario \citep{2017PhRvD..95l4046G,PhysRevD.100.043027,2019PhRvD.100d1301G,2019arXiv191104424D}. 

It is plausible that globular clusters (GCs) started life as dense gas clouds, which are referred to as protoglobular or primordial clusters. They underwent prolonged star formation early in their lifetimes \citep{Gratton_2012A&ARv..20...50G}. A huge gas reservoir, comparable to or higher in mass than that of its stellar component, is available for accretion immediately after the first stellar formation event. 
Feedback processes from stellar evolution in star-forming regions are believed to clear away the surrounding gas (e.g., \citealt{Voss_2010A&A...520A..51V,Galvan-Madrid_2013ApJ...779..121G,Krumholz_2014prpl.conf..243K}). The effectiveness of the process depends on the compactness  \citep{Krause_2012A&A...546L...5K,Krause_2016A&A...587A..53K,Silich_2017MNRAS.465.1375S,Silich_2018MNRAS.478.5112S}, that is, the mass over half-mass radius of the cluster,
\begin{equation}\label{eq:C}
C = \frac{M_\text{total}}{10^5 M_\odot}\left( \frac{r_\text{hm}}{pc}\right)^{-1}.
\end{equation} 
For sufficiently compact clouds, feedback processes are expected to become ineffective because they are proportional to the total mass. The gravitational binding energy is proportional to the square of it. 

Proposals regarding the precise mechanism for gas depletion in primordial GCs abound in the literature, but a general consensus has not been achieved \citep{Spergel_1991Natur.352..221S,Thoul_2000LIACo..35..567T,Fender_2005MNRAS.360.1085F,Moore_2011ApJ...728...81M,Herwig_2012ApJ...757..132H,Krumholz_2014prpl.conf..243K,Krause_2016A&A...587A..53K,Silich_2017MNRAS.465.1375S,Silich_2018MNRAS.478.5112S,Marks_2008MNRAS.386.2047M,Kruijssen_2012MNRAS.426.3008K,DErcole_2008MNRAS.391..825D,Conroy_2012ApJ...758...21C,Renzini_2015MNRAS.454.4197R,Bagetakos_2011AJ....141...23B,Krause_2013A&A...550A..49K,Jaskot_2011ApJ...729...28J,Fierlinger_2016MNRAS.456..710F,Yadav_2017MNRAS.465.1720Y,Krause_2012A&A...546L...5K}. \cite{Leigh_2013MNRAS.429.2997L} proposed that in any cluster that is able to form massive stars, the primordial gas is depleted exactly due to the accretion onto BHs. They find that accreting BHs can deplete the whole gas reservoir within as short a time as $10Myr$.

Here, we do not investigate this specific scenario and do not focus on the gas reservoir, but on the effect of accretion onto the BHs of the cluster. 
\cite{Calura_2015ApJ...814L..14C} estimated that within $\sim 14Myr$ the gas is depleted {by} $99\%$  by star formation feedback processes in a primordial cluster with an initial total mass $\text{of about } 10^7M_\odot$. 
Typical globular clusters (GCs) should form $\text{about } 10^2-10^3$ BHs within $\text{about } 3 Myr$  \citep{2013ApJ...763L..15M,Morscher_2015ApJ...800....9M} and could retain most of them initially if their natal kicks are sufficiently low (see \citealt{Wong_2012} and references therein). Additional observational evidence suggests that a BH subcluster may be retained even to present-day GCs \citep{sedda2019moccasurvey,Abbate:2019wzu}.

We assume in our analysis an initial population of $500$ BHs, segregated in the core of the cluster \citep{Spitzer_1987degc,2014MNRAS.441..919L}, and that the primordial gas is depleted by $99\%$ {within time $t_f= 10 Myr$,} following an exponential law,
\begin{equation}\label{eq:rho_gas_t}
\rho_\text{gas} = \rho_\text{gas}(0) e^{-t/\tau},\quad 
\tau = \frac{t_f}{2 \ln(10)}.
\end{equation}
Therefore our analysis applies to any type of depletion mechanism as long as it does not proceed faster than an exponential law and the loss is approximately uniform in the core. To the exponential gas loss, we add the loss by accretion onto the BHs.

In the next section we briefly describe our model, and in section \ref{sec:results} we present the results of our analysis. In the final section we discuss our conclusions. 

\section{Model}\label{sec:model}

We considered $N_\bullet = 500$ BHs moving inside a fixed external potential with gaseous and stellar components. This analysis focuses solely on the effect of accretion onto the BH mass function, and in addition, of accretion within a short timescale $t_f = 10Myr$. We do not study the general effect on the cluster or BH kinematics. For simplicity and with the aim to provide only statistical estimates, 
we therefore assumed a spherical distribution of BHs at any $t$, that is, we did not follow the angular changes of their orbital planes (the BH subsystem dynamics is dominated by the gaseous and stellar components on the timescales considered). We further neglected close encounters. We discuss this further in our conclusions. We intend to include and study both of these  effects in a separate more detailed work that will include the estimation of merger rates.  

We assumed that the fixed external potential is generated by Plummer density profiles for stellar and gaseous components,
\begin{align}
\label{eq:rho_stars}
\rho_\star (r) &= \frac{3\epsilon M_\text{total}}{4\pi a^3}\left( 1 + \frac{r^2}{a^2}\right)^{-5/2},
\\
\label{eq:rho_gas}
\rho_\text{gas} (r,t) &= 
\frac{3 (1-\epsilon) M_\text{total} }{4\pi a^3}\left( 1 + \frac{r^2}{a^2}\right)^{-5/2}
e^{-t/\tau}, 
\end{align}
where $\epsilon$ is the stellar formation efficiency, and $\tau$, given in Eq. (\ref{eq:rho_gas_t}), corresponds to $99\%$ gas depletion by the final time $t_f$. 
Following \cite{Silich_2018MNRAS.478.5112S}, we assumed that the gas is dominated by turbulence \citep{2015ApJ...806...35J,2017ApJ...836...80E}, in which case the equation of state is $P_\text{gas} = \rho_\text{gas} \sigma_\text{gas}^2$.

The BHs were chosen from a Salpeter  initial mass function \citep{2019ApJ...878L...1P} with a cutoff
\begin{equation}
f(m_\bullet) = 
\left\lbrace 
\begin{array}{ll}
\displaystyle A m_\bullet^{-2.35} &, \, 5M_\odot \leq m_\bullet \leq 50M_\odot \\[2ex]
0 &,\, m_\bullet < 5M_\odot\;\text{or}\;    m_\bullet > 50M_\odot
\end{array}
\right.,
\end{equation}  
with $A = 1.35/(5^{-1.35}-50^{-1.35})$.
They were all assumed to be initially bound inside the core radius
$
r_\text{c} = 0.64 a
$
of the initial Plummer sphere. The initial radial distribution and velocities of the BHs were chosen randomly from a Plummer distribution with the same softening radius $a$ of Eqs. (\ref{eq:rho_stars}) and (\ref{eq:rho_gas}). Their initial angular distribution was spherical.

Hot-type accretion, whose typical representative is Bondi-Hoyle accretion, is appropriate in the dense gaseous environment we examined, where the speed of sound is greater than or on the order of the accretor velocity \citep{Merritt_book}. 
The spherically symmetric accretion rate for a given cross section $\pi R_\text{acc}^2$ may be written as \citep{1985Sci...230.1269F}
\begin{equation}\label{eq:dm}
\dot{m}_\bullet =       \pi \rho_\text{gas} v_\text{rel} R_\text{acc}^2,
\end{equation}
where $m_\bullet$ is the mass of the BH that accretes the gas with relative velocity 
\begin{equation}
v_\text{rel} = \sqrt{v^2 + c_s^2},
\end{equation} 
and $v$ is the BH velocity with respect to the center of mass of the cluster, and $c_s$ is the speed of sound of the gas.

As \cite{2001MNRAS.324..573B} pointed out, choosing the proper accretion radius inside a gaseous stellar cluster depends on the relative amount of gas and on the radial position of the accretor. When the gas dominates the cluster potential, the accretion rates are given by a tidal-lobe accretion radius \citep{1971ARA&A...9..183P},
\begin{equation}
R_\text{tid}(r_i) \sim 0.5 \left( \frac{m_\bullet}{M(r<r_i)} \right)^{1/3} r_i,
\end{equation}
where $M(r<r_i)$ is the total mass of the cluster within the radial position $r_i$ of the $i$th BH. When the gas has been sufficiently depleted so that stars dominate the potential, the appropriate accretion radius is the Bondi – Hoyle radius,
\begin{equation}
R_\text{B} = \frac{2 G m_\bullet}{v_\text{rel}^2}.
\end{equation}
Following \cite{2001MNRAS.324..573B}, and with the intention to provide a minimum estimate of the BH mass growth, we chose the accretion radius to be the smaller of the two,
\begin{equation}\label{eq:R_acc}
R_\text{acc}(t) = \min\{ R_\text{B},R_\text{tid} \},
\end{equation}
at any instant of time.

We accounted in the analysis for the dynamical friction generated by the stellar component according to Chandrasekhar's formula
\begin{equation}\label{eq:df_chandra}
F_{\text{df},\star} = \frac{4 \pi G^2 m_\bullet^2}{v^2}\rho_\star \ln\Lambda
\left\lbrace \text{Erf}(\textstyle\frac{v}{\sigma \sqrt{2}}) - \frac{2}{\sqrt{\pi} }  \frac{v}{\sigma \sqrt{2}}e^{-\frac{v^2}{2\sigma }} \right\rbrace ,
\end{equation}
where $v$ is the velocity of the BH, and the Coulomb logarithm is $\ln\Lambda = \ln (b_\text{max}/b_\text{min})$ with
\begin{equation}
b_\text{max} = r_\text{hm},\quad
b_\text{min} = \frac{G m_\bullet}{3v_\text{rel}^2}.
\end{equation}

\cite{2011MNRAS.416.3177L,2014A&A...561A..84L} reported that the gaseous dynamical friction on an accretor is 
\begin{equation}\label{eq:a_acc}
\bm{a}_\text{acc} = -\frac{\dot{m}}{m}\bm{v}
\end{equation}
for both subsonic and supersonic accretors. This result seems to agree with the earlier calculation of \cite{Hadjidemetriou_1963Icar....2..440H}. It may be understood simply as manifesting angular momentum preservation (see also \cite{2019A&A...621L...1R}).
\cite{2011MNRAS.416.3177L,2014A&A...561A..84L} argued that Eq. (\ref{eq:a_acc}) encompasses the entire gaseous dynamical friction. In contrast, \cite{2009ApJ...696.1798T} added to this term the gaseous dynamical friction formula proposed by \cite{1999ApJ...513..252O} (see also   \citealt{1996A&A...311..817R,2001MNRAS.322...67S,Kim_2009,2013JApA...34..207I,Antoni_2019}). This issue seems unresolved. We adopted the former view (only the term in Eq. (\ref{eq:a_acc}), neglecting Ostriker's term) because it generates the least gaseous dynamical friction. We wish to provide minimum estimates of BH growth, and higher dynamical friction causes BHs to sink deeper into the center where the gas density is higher, leading to more intense BH growth. Including Ostriker's formula would only lead to results that support our conclusions more strongly. We can also report that when Ostriker's formula is used instead of Eq. (\ref{eq:a_acc}), we obtained similar results.

Finally, we describe the equations of motion of the BH-subcluster. 
As we noted, the BHs were assumed to be initially spherically distributed, and we are interested only in the amount of gas that they can accrete within $10Myr$. Therefore we did not follow the angular redistribution of their orbital planes and also assumed that the background of stars and gas is spatially fixed, as in Eqs. (\ref{eq:rho_stars})-(\ref{eq:rho_gas}). We further assumed that the gas is depleted by $99\%$ uniformly and exponentially within $t_f = 10Myr$ due to any process. The BHs accrete gas according to Eq. (\ref{eq:dm}) and are subject to dynamical friction generated by stars, Eq. (\ref{eq:df_chandra}), as well as gas, Eq. (\ref{eq:a_acc}). The system of equations that we solved numerically for $i=1,\ldots,N_\text{BH}$ was therefore
\begin{align}
\frac{d r_i}{dt} &= v_{r,i} ,\\
\frac{d \phi_i}{dt} &= \frac{v_{\phi,i}}{r_i} ,\\
\frac{d v_{r,i}}{dt} &= -\frac{v_{\phi,i}^2}{r_i} -\frac{G M_\text{enc}}{r_i^2} - (a_{\text{df},\star,i} + a_{\text{acc},i})\frac{v_{r,i}}{v_i},\\
\frac{d v_{\phi,i}}{dt} &= -\frac{v_{r,i} v_{\phi,i}}{r_i} - (a_{\text{df},\star,i} + a_{\text{acc},i})\frac{v_{\phi,i}}{v_i}, \\
\frac{dm_i}{dt} &= \pi \rho_\text{gas}(r_i) v_\text{rel}\left(v_i,c_s(r_i)\right) R_\text{acc}\left(v_i,c_s(r_i)\right)^2.
\end{align}
The total mass within $r_i$, $M_\text{enc} = M(r<r_i(t))$, includes the enclosed mass of field stars, of the remaining enclosed gas at $t,$ and the enclosed mass within $r_i$ of the BH population.

\section{Results}\label{sec:results}

\begin{figure}[tb]
        \begin{center}
                \includegraphics[scale = 0.5]{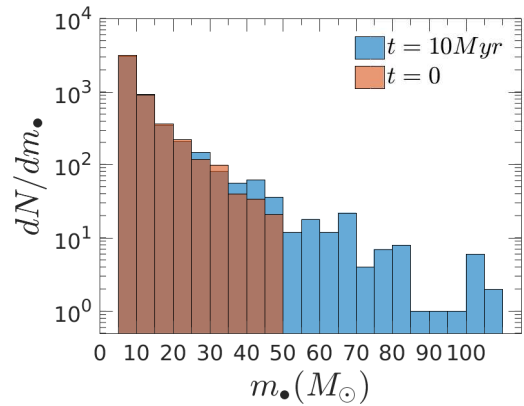}  
                \caption{Total BH mass distribution of $\text{ten}$ samples of $N_\bullet = 500$ BHs, with an initial mass cutoff at $50M_\odot$, bound in the core of a primordial gaseous stellar cluster with $r_\text{hm} = 1pc$. We assumed the gas to be exponentially depleted by $99\%$ within $10Myr$. }
                \label{fig:m_histo}
        \end{center} 
\end{figure}

\begin{figure}[tb]
        \begin{center}
                \includegraphics[scale = 0.5]{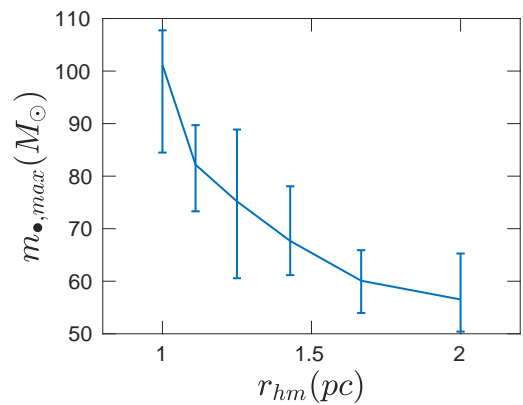}  
                \caption{{Maximum} BH mass achieved after gas accretion of an initial population of $N_\bullet = 500$BHs, with initial mass cutoff at $50M_\odot$, with respect to the half-mass radius $r_{hm}$ of the stellar cluster for $\text{ten}$ samples of the BH population at each $r_{hm}$. The vertical lines represent the range of maximum mass values achieved in all samples for a certain $r_{hm}$. The gas is assumed to be exponentially depleted by $99\%$ within $t_f = 10Myr$.}
                \label{fig:m_max}
        \end{center} 
\end{figure}

We performed the numerical analysis for an initial total cluster mass $M_\text{tot} = 10^6 M_\odot$, stellar formation efficiency $\epsilon=0.3$, and several initial compactness values $C_\text{ini}(t=0) = \{5-10\}$, corresponding to half-mass radii $r_\text{hm} = \{2-1\} pc$. For each half-mass radius we performed $\text{ten}$ simulations with different initial conditions. We also assumed a depletion time $t_f=10Myr$. 

In Figure \ref{fig:m_histo} we show the final BH mass function in the case of a half-mass radius $r_\text{hm} = 1pc$, that is, an initial compactness $C_\text{ini}=10$, which results in a final compactness $C_\text{fin} = 3$. It is evident that about $2\%$ of the BHs fill in the theoretical BH mass gap. In addition, the BH mass function increases for mass values higher than $\gtrsim 35M_\odot$. This is also true for any $r_\text{hm} \leq 1.25pc$.
In Figure \ref{fig:m_max} we draw the maximum BH mass that is achieved after accretion with respect to the half-mass radius of the cluster.

We report that the probability for a BH to exceed the initial cutoff of $50M_\bullet$ is
\begin{equation}
P(m_\bullet > 50M_\odot) \simeq 1-2\% \,,\quad
\text{for}\; r_\text{hm} = \{1-2\}pc,
\end{equation}
where the lower half-mass radius corresponds to the upper probability value.
The total accreted gas mass lies in the range $\simeq \{100-900\}M_\odot$ (with an initial mass $5400M_\odot$ of the BH population).

\begin{table}
        \begin{center}
                \begin{tabular}{  c | c }
                        \toprule
                        $m_\bullet (M_\odot)$ & $a_\text{BBH}(10^{-3}pc)$
                        \\
                        \midrule
                        $\{108,\,45\}$
                        &
                        $0.1$
                        \\
                        $\{82 ,\, 64\}$
                        &
                        $1.6$
                        \\
                        $\{73 ,\, 51\}$
                        &
                        $0.8$
                        \\
                        $\{90 ,\, 63\}$
                        &
                        $0.1$
                        \\
                        $\{78 ,\, 75\}$
                        &
                        $0.9$
                        \\
                        $\{84 ,\, 68\}$
                        &
                        $1.6$
                        \\
                        \bottomrule
                \end{tabular}
        \end{center}
        \caption{In five out of the ten samples with $r_\text{hm} = 1pc$, six massive BBHs formed within $10Myr$ because they sank to the center of the cluster ($\lesssim 10^{-3}pc$). We assumed a population of $N_\bullet = 500$ BHs, with different initial conditions in each sample, immersed inside the core of the same gaseous primordial cluster. The left column denotes the masses of each BBH member, and the right column lists the semi-major axis of each BBH.  \label{tab:m_BBH}
        }
\end{table}

We remark that the more massive BHs sink deeper into the center $r \lesssim 10^{-3}pc$, where it is most probable that they form hard massive BBHs. About half of the samples at each $r_\text{hm}$ form a BBH by this process. In particular, for $r_\text{hm} = 1pc$, six hard BBHs were formed in five out of the ten samples, as shown in Table \ref{tab:m_BBH}. The BBHs will continue to form for $t>10Myr$ by dynamical processes.

\section{Conclusions}

We propose that in the early life of dense stellar clusters a rapid accretion process operates. The BHs that have been generated by the more massive stars, most of which are therefore segregated in the core, rapidly accrete primordial gas --that is also accumulated in the core-- within $\sim10Myr$ before it becomes depleted by stellar formation feedback processes. The amount of accreted gas depends sensitively on the mass and compactness of the cluster and on the accretion timescale. 

The BH (and BBH) initial mass function are shifted toward higher values. For sufficiently compact clusters the BH mass limit of $50M_\bullet$  predicted by supernova theory is exceeded. 
We estimate that $1-2\%$ of the initial BH population exceeds the theoretical limit for an initial cluster mass $M_\text{tot}=10^6M_\odot$, half-mass radii $r_\text{hm}=\{1,2\}pc$, depletion time $t_f=10Myr,$ and stellar formation efficiency $\epsilon = 0.3$. 

In addition, the BH mass growth together with dynamical friction causes the more massive BHs to sink deeper into the center of the cluster. There, it is more probable that the BHs attains mutual binding energies that are higher than each individual binding energy with the cluster and thus forms hard massive BBHs, as listed in Table \ref{tab:m_BBH}. 
This process operates rapidly in the very early life of the cluster, therefore it adds {to the} dynamical channel of BBH formation and mergers \citep{Abbott_2016ApJ...818L..22A}, which will continue to operate during the whole life of the cluster (e.g., see \citealt{2017MNRAS.469.4665P,2018MNRAS.480.5645H}). {Thus, an} observation  by LIGO of BBH mergers with member masses above $50M_\bullet$ {is to be} anticipated.

We assumed an initial BH population bound inside the core of the cluster {neglecting} close encounters. These are not expected to have a strong effect on our results because their timescale are {relatively} long. The recoil mechanism of BBH-BH three-body encounters, which ejects BBHs out of the cluster \citep{Sigurdsson_1993Natur}, operates on $\text{gigayear}\text{}$ timescales, but our proposed mechanism works on $\text{megayear}$ timescales. In addition, it has become evident during the past decade \citep{2007Natur.445..183M,2011ApJ...734...79B,2010ApJ...721..323S,2011MNRAS.410.1655M,2012Natur.490...71S} that recoil is not effective in sufficiently dense GCs, which contain a population ($\lesssim 1000$) of BHs and BBHs in their core (e.g., \citealt{Morscher_2015ApJ...800....9M} and references therein). 

More sophisticated simulations may be required in order to confirm with higher confidence and detail the effects of the rapid accretion process we proposed here and to provide a better estimation of the resulting BH mass function for a wider range of parameter values $\{M_\text{tot}, r_\text{hm}, t_f, \epsilon \}$. 
Nevertheless, our current analysis strongly supports the idea that the BH upper mass gap can be populated by rapid gas accretion onto the BHs of dense primordial stellar clusters.

\bibliography{2019_BH_massgap_arXiv}
\bibliographystyle{aa}

\end{document}